\documentclass[twocolumn]{aastex62}
\usepackage{textcomp}
\usepackage{amsmath,amstext}
\usepackage{mathptmx}
\usepackage{txfonts}
\usepackage{bm}
\usepackage[T1]{fontenc}
\usepackage{ae,aecompl}
\usepackage{hyperref}

\graphicspath{{./}{figures/}}

\received{October 19, 2018}
\revised{December 14, 2018}
\accepted{December 16, 2018}

\shorttitle{LATE DELIVERY OF NITROGEN TO EARTH}
\shortauthors{Chen et al.}

\begin{document}

\title{Late delivery of nitrogen to Earth}

\correspondingauthor{Cheng	Chen}
\email{chenc21@unlv.nevada.edu}

\author{Cheng Chen}
\author{Jeremy L. Smallwood}
\author{Rebecca G. Martin}
\author{Mario Livio}

\affiliation{Department of Physics and Astronomy, University of Nevada, Las Vegas, 4505 S. Maryland Pkwy.Las, Box 454002, Vegas, NV 89154, USA}

\begin{abstract}
Atmospheric nitrogen may be a necessary ingredient for the habitability of a planet since its presence helps to prevent water loss from a planet. The present day nitrogen isotopic ratio, $^{15}$N/$^{14}$N, in the Earth's atmosphere is a combination of the primitive Earth's ratio and the ratio that might have been delivered in comets and asteroids. Asteroids have a  nitrogen isotopic ratio that is close to the Earth's. This indicates either a similar formation environment to the Earth or that the main source of nitrogen was delivery by asteroids. However, according to geological records, the Earth's atmosphere could have been enriched in $^{15}$N during the Archean era. Comets have higher a $^{15}$N/$^{14}$N ratio than the current atmosphere of the Earth and we find that about $5\%$ $\sim$ $10\%$ of nitrogen in the atmosphere of the Earth may have been delivered by comets to explain the current Earth's atmosphere or the enriched $^{15}$N Earth's atmosphere. We model the evolution of the radii of the snow lines of molecular nitrogen and ammonia in a protoplanetary disk and find that both have radii that put them farther from the Sun than the main asteroid belt. With an analytic secular resonance model and N--body simulations we find that the $\nu_8$ apsidal precession secular resonance with Neptune, which is located in the Kuiper belt, is a likely origin for the nitrogen-delivering comets that impact the Earth.  
%Due to the outward migration of Uranus and Neptune after their formation, the $\nu_8$ resonance swept through the belt driving comets inwards to collide with the Earth and deliver the nitrogen.

\end{abstract}


\section{Introduction} 
\label{sec:intro}
Nitrogen may be an essential part of an atmosphere for a planet to be habitable \citep[e.g.][]{Stucken2016}. Nitrogen is a component of chemical compounds that are necessary for growth and reproduction of all animal and plant life on Earth. It is found in amino acids that make up proteins as well as in nitrogenous bases in nucleic acids that store, transcribe and translate the genetic code.  Furthermore, nitrogen influences the surface temperature and water content of the atmosphere, both of which affect planet habitability.  
The nitrogen content in the atmosphere plays a significant role in the rate of water--loss from a planet \citep[e.g.][]{Wordsworth2013b,Wordsworth2014}. If the nitrogen level is too low, water loss may be very efficient for all surface temperatures.

Nitrogen is the most abundant molecule in the atmosphere of the Earth, comprising about $78\%$ of our atmosphere by volume, but still, its origin is not well understood. The reservoir of nitrogen in the crust and the mantle of the Earth currently is thought to be between 0.4 and 7 times the present day atmospheric amount \citep[e.g.][]{Marty1995, Halliday2013,Johnson2015}. Isotopic measurements suggest that most of the current atmospheric nitrogen is from the inner solar system \citep{Furi2015}. The relative abundance of the two isotopes of nitrogen $^{15}$N/$^{14}$N in the Earth is similar to Venus, Mars and primitive meteorites such as chondrites \citep[e.g.][]{Hutsemekers2009,Marty2012,Harries2015,Bergin2015}.

In the case of the Earth we know that most of the nitrogen was accreted locally. However, this may not be universally the case. Therefore, a study of potential habitability (assuming that nitrogen is a necessary ingredient) needs to explore other potential delivery mechanisms.

The isotopic ratio of nitrogen in the present day atmosphere is measured to be $^{15}$N/$^{14}$N = $3.676 \times 10^{-3}$ \citep{JUNK1958}. However, the Earth's mantle is relatively depleted in  $^{15}$N compared to the Earth's atmosphere. We compare the ratios with 
\begin{equation}
\rm \delta^{15}N = \left(\frac{(^{15}N/^{14}N)_{mantle}}{(^{15}N/^{14}N)_{atmosphere}}-1\right) \times 1,000.
\end{equation}

In the Earth's mantle, $\delta^{15}$N  is estimated to be in the range -5 \textperthousand \, to -30 \textperthousand \citep{Cartigny2013}. 
Thus, the isotopic ratio of nitrogen in the Earth's mantle is in the range between $^{15}$N/$^{14}$N = $3.566 \times 10^{-3}$ to $3.658 \times 10^{-3}$  \citep{Cartigny1998,Cartigny2013,javoy1986,Jia2004,Marty1999,Murty1997}. Moreover, according to the record of Archean sedimentary rocks and crustal hydrothermal systems, the Earth  may have had a $^{15}$N-enriched atmosphere with $\delta^{15}$N$=30$ \textperthousand \ and $^{15}$N/$^{14}$N = $3.786 \times 10^{-3}$ in Archean eon \citep{Jia2004}, 4 to 2.5 billion years ago. 
This high estimate has later been questioned because it was based on the analysis of metamorphosed crustal rocks \citep{PAPINEAU2005,ADER2006,STUEKEN2015}. Different geological records of less metamorphosed Archean sedimentary rocks suggest a lower value $\delta^{15}$N in the range 0\textperthousand \ to 15\textperthousand  \citep{sano1990,PINTI2001} and more recent estimates based on the analysis of fluid inclusions having trapped Archean seawater concluded that the N isotope composition of the Archean atmosphere was close to 0\textperthousand \citep{Marty2013,AVICE2018}.
In our calculations (see Sec.~\ref{nitrogen}), we consider both a $^{15}$N-rich atmosphere and the present day atmosphere.

Several possible mechanisms may explain a high value of $^{15}$N/$^{14}$N during the Archean era, such as hydrodynamic escape of light isotopes or heterogeneous accretion of the Earth that forms from material with $^{15}$N-depleted and then accretes a late veneer composed of chondritic material and a minor contribution of $^{15}$N-enriched comets. However, these mechanisms are difficult to quantify \citep{Cartigny2013,JAVOY1998,Marty2012}.  It is most likely that to reach the current nitrogen isotopic ratio of the Earth's atmosphere or the $^{15}$N-enriched atmosphere in the Archean era, there must have been some comets delivered to the Earth after the Earth formed and differentiated \citep{Jia2004,Rice2018}.

Enstatite chondrites  formed in the inner solar system and could be the primary ingredient to form terrestrial planets \citep{Wasson1988}. These chondrites can carry nitrogen in osbornite, sinoite and nierite \citep{Rubin2009}. On the other hand, nitrogen was probably delivered to the Earth by comets in the form of ammonia ices or simple organic compounds such as HCN. In laboratory studies, molecular nitrogen can be trapped and released by amorphous ice \citep{Barnun1988}. Although in the case of Halley's comet, only a small amount of $\rm N_{2}^{+}$ was detected due to the enrichment of {\rm CO} over ${\rm N_{2}}$ in amorphous ice \citep{Notesco1996}. The abundance of nitrogen in the surface layer of a comet measured by a satellite does not necessarily represent the ratio in its nucleus \citep{Prialnik1992}.  In the case of Halley's comet, most of the nitrogen is in the form of compounds \citep{Krankowsky1991, Geiss1987,Geiss1988, Meier1994}. The amount of nitrogen that can be trapped in ice at temperatures of around $30 \,\rm K$ is five orders of magnitude higher than in ice at a temperature of $100\,\rm K$ \citep{Barnun1988}. However, in comets, noble gases have about an equal abundance as molecular nitrogen. If these were delivered to the Earth, the Earth would have had an overabundance of noble  gases \citep{Owen1995ii,Marty2017}. Nevertheless, a recent analysis from the Rosetta Orbiter Spectrometer for Ion and Neutral Analysis (ROSINA) revealed that the relative abundance of Ar/$\rm N_2$ inside the icy nucleus of the 67P/Churyumov-Gerasimenko is $ 9.1 \pm 0.3 \times 10^{-3}$ \citep{Balsigere2015}. Therefore, comets may not contain the same amount of noble gases as they do nitrogen. Moreover, the xenon isotopic composition shows a deficit in heavy xenon isotopes from ROSINA. By considering the origin of a primordial atmospheric xenon component is a mixture of chondritic and cometary xenon, the present Earth's atmosphere contains 22$\pm$5\% cometary xenon \citep{Marty2017}.

The nitrogen isotopic ratio could change in the outer regions of the solar system where comets formed at extremely low temperature. Large enhancements of the isotopic ratio $^{15}$N/$^{14}$N can occur in cold, dense gas where CO is frozen out. Particularly, if  the temperature is around $7 - 10 \,$K, then the upper layers of the grains can be enhanced in $^{15}$N by up to an order of magnitude \citep{Rodgers2008}. This mechanism introduces an uncertain factor in the measured $^{15}$N/$^{14}$N isotopic ratio in comets. Some comets have an anomalous nitrogen isotopic ratio that is about twice the terrestrial ratio \citep{manfroid2009}.

%Understanding the evolution of the Earth's nitrogen can help us to study the habitability in exoplanets. To investigate biological evidence beyond the Earth, Raman scattering can help to identify atmospheric properties of exoplanets by imprinting in the reflected light and the geometric albedo of exoplanets, even molecular nitrogen, which is not able to be identified in the optical wavelength range \citep{Oklopcic2016}. 

In this work we explore the possibility of late delivery for the $^{15}$N-enriched nitrogen in the atmosphere of the Earth. In Section~\ref{disk} we examine the evolution of a protoplanetary disk in which planetesimals later form. We calculate the location of the ammonia and nitrogen snow lines -- the radius outside of which each would have condensed out of the gas and would have been incorporated into the solid bodies. While there is evidence that most of the nitrogen was delivered locally it is still interesting to explore whether some of it could be delivered at late times, via comets or asteroids. In Section~\ref{nitrogen} we consider what isotopic nitrogen ratios tell us about how much of the nitrogen could have been delivered by comets and find that it could have been a significant fraction. In Section~\ref{asteroid} we examine possible delivery mechanisms to Earth for the relevant solid bodies. In particular, we show how comets are perturbed by mean motion and secular resonances with the planets in the solar system and we use $N$-body simulations to show that resonances can play a role in the delivery of  comets to the Earth. Finally, we draw our conclusions in Section~\ref{con} .

\section{Protoplanetary disk evolution}
\label{disk}

The temperature of a protoplanetary disk generally decreases with distance from the central star.  A snow line marks the radial distance  from the star outside of which the temperature drops below the condensation temperature of a particular compound. The condensation temperature for hydrated ammonia is about $T_{\rm snow,NH_3} =131\,\rm K$, and for hydrated nitrogen is about $T_{\rm snow,N_2}=58\,\rm K$ \citep{Lodders2003}. In this Section we consider the evolution of these snow lines in a protoplanetary disk.  The snow lines determine where ammonia and nitrogen were  incorporated in a solid form into planetesimals that formed there. 

Material in the disk interacts through viscosity that is thought to be driven by the magnetorotational instability (MRI) that drives turbulence in the disk \citep{BH1991}.  We first consider the snow line radii in a steady-state fully turbulent disk model. However, protoplanetary disks are thought to contain a region of low turbulence at the disk midplane, known as a "dead zone", where the disk is not sufficiently ionized for the magnetorotational instability to operate \citep[e.g.,][]{Gammie1996, Gammie1998}. Thus, we also consider how the presence of a dead zone affects the evolution of the snow lines. We follow \citet{MartinandLivio2014} and derive analytic solutions for the snow line radii in each case.

\subsection{A Fully Turbulent MRI Disk}

%To evaluate the snow line radii of nitrogen, $\rm N_{2}$, and ammonia, $\rm NH_{3}$,  We  consider the temperature balance in the disc with a cooling term including black body radiation and the heating terms including the viscous heating of disc and the irradiation from a host star. 

\begin{figure}
  \centering
    \includegraphics[width=8.3cm]{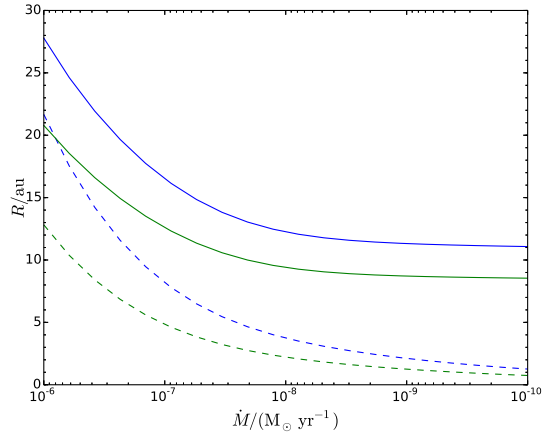}
     \caption{Nitrogen (blue) and ammonia (green) snow lines as a function of accretion rate in a steady state disk with $M_{_\star}=\rm{M_{\odot}}$, $T_{_\star}=4000\, \rm K,\, R_{_\star}=3\, \rm R_{\odot}$, $T_{\rm N_{2},snow}=58\, \rm K$ and $T_{\rm NH_{3},snow}=131\,\rm K.$ The dashed lines show the snow lines with the fully MRI turbulent disk with $\alpha_{\rm m}$=0.01. The solid lines represent the snow lines in a disk in a self-gravitating dead zone. } 
     \label{sla}
\end{figure}
    
Material in an accretion disk orbits the central star of mass, $M_{_\star}$, with Keplerian velocity at radius $R$  with angular velocity $\Omega=\sqrt[]{GM_{_\star}/R^3}$ \citep{Pringle1981}. The viscosity in a fully MRI turbulent disk is 
\begin{equation}
\nu = \alpha_{\rm m}\frac{c_{\rm s}^{2}}{\Omega},
\end{equation}
where $\alpha_{\rm m}$ is the viscosity parameter \citep{SS1973} and the sound speed is $c_{\rm s}=\sqrt[]{{\cal R}T_{\rm c }/\mu}$, where $\cal R$ is the gas constant, $\mu$ is the gas mean molecular weight and $T_{c}$ is the mid-plane temperature. 
With a constant accretion rate, $\dot{M}$, through all radii, the surface density of a steady state disk is 
\begin{equation}
\Sigma = \frac{\dot{M}}{3\pi\nu}
\end{equation}
\citep{Pringle1981}. The surface temperature is determined through the balance of heating from viscous effects and irradiation from the central star
\begin{equation}
\sigma T_{\rm e}^4 = \frac{9}{8}\frac{\dot{M}}{3\pi}\Omega^2+\sigma T_{\rm irr}^4
\label{sdd}
\end{equation}
\citep[e.g.][]{Cannizzo1993,Pringleetal1986}, where $\sigma$ is the Stefan-Boltzmann constant and the irradiation temperature is 
\begin{equation}
T_{\rm irr}=\left(\frac{2}{3 \pi}\right)^{\frac{1}{4}}\left(\frac{R_{_\star}}{R}\right)^{\frac{3}{4}}T_{_\star}
\label{Trr}
\end{equation}
\citep{Chiang1997}, where $T_{_\star}$ is the temperature and $R_{_\star}$ is the radius of the star. The mid-plane temperature is related to the surface temperature through
\begin{equation}
T_{\rm c}^4=\tau T_{\rm e}^4,
\label{Tc}
\end{equation}
where the optical depth, $\tau$,  is given by
\begin{equation}
\tau=\frac{3}{8}\kappa \frac{\Sigma}{2}
\label{opacity}
\end{equation}
and the opacity is $\kappa=aT_{\rm c}^{\rm b}$. In our model, we take a = 0.0001 and b = 2.1 since dust dominates the absorption properties of matter in the disk \citep{Armitage2001}. 

The snow line radius in the disk is found by solving $T_{\rm c}=T_{\rm snow}$. In our model, we assume $ M_{_\star}=1\, {\rm M_{\odot}}$, $R_{_\star}=3\, {\rm R_{\odot}}$ and  $T_{_\star}=\,4000 {\rm K}$. There remains some uncertainty in the value of the viscosity parameter \citep[e.g.][]{King2007}. Observations of FU Ori show $\alpha_{\rm m}\approx 0.01$ \citep{Zhu2007} while from observations of X-ray binaries and cataclysmic variables, it is estimated that $\alpha_{\rm m} \approx 0.1 \sim 0.4$ \citep{King2007}. Since we consider young stellar objects, we assume that  $\alpha_{\rm m} =  0.01$ in a fully turbulent MRI disk. 

 We adapt the approximate analytical steady state solution of the CO snow line in \citet{MartinandLivio2014} and find analytic fits to the nitrogen snow line radius
\begin{equation} 
\begin{split}
R_{\rm N_{2}, snow} \approx 3.6\left(\frac{\alpha_{\rm m}}{0.01}\right)^{-\frac{2}{9}}\left(\frac{M_{_\star}}{\rm M_{\odot}} \right)^{\frac{1}{9}}
\left(\frac{\dot{M}}{10^{-8}\, \rm M_{\odot}\, yr^{-1}} \right)^{\frac{2}{9}} \\
\times\left(\frac{T_{\rm N_{2},snow}}{\rm 58K} \right)^{-0.64} \left( \frac{R_{_\star}}{3\,\rm R_{\odot}} \right)^{\frac{2}{3}}  \left(\frac{T_{_\star}}{4000\,\rm K} \right)^{\frac{8}{9}}\, \rm au
\end{split}
\label{r1}
\end{equation}
and the ammonia snow line radius
\begin{equation}
\begin{split}
R_{\rm NH_{3}, snow} \approx 2.1\left(\frac{\alpha_{\rm m}}{0.01}\right)^{-\frac{2}{9}}\left(\frac{M_{_\star}}{\rm M_{\odot}} \right)^{\frac{1}{9}}
\left(\frac{\dot{M}}{10^{-8}\, \rm M_{\odot}\, yr^{-1}} \right)^{\frac{2}{9}} \\
\times\left(\frac{T_{\rm NH_{3}, snow}}{\rm 131K} \right)^{-0.64} \left( \frac{R_{_\star}}{3\,\rm R_{\odot}} \right)^{\frac{2}{3}}  \left(\frac{T_{_\star}}{\rm 4000\ K} \right)^{\frac{8}{9}}\, \rm au.
\end{split}
\label{r2}
\end{equation}
These are valid for relatively low accretion rates, $\dot{M}\lesssim 10^{-8}\, \rm M_{\odot}\, yr^{-1}$, where irradiation dominates the viscous heating term. 

The numerical values of the snow line radii in equations~(\ref{r1}) and~(\ref{r2}), for the standard parameters chosen, are uncomfortably small. The ammonia snow line lies inside of the main asteroid belt while the nitrogen snow line lies within it.  Fig.~\ref{sla} shows the full numerical solution for the snow line radii for nitrogen and ammonia in a fully turbulent disk as a function of the steady-state accretion rate (dashed lines) found by solving $T_{\rm c}=T_{\rm snow}$ along with equations~(\ref{sdd}) to~(\ref{Tc}). This exposes the problem with fully turbulent disks, in that the snow line moves in too close to the host star during the low accretion rate phase when the disk is near the end of its lifetime \citep{GaraudandLin2007, Okaetal2011, MartinandLubow2011, MartinandLivio2014}. This contradicts  observations of the composition of asteroids in the asteroid belt \citep[e.g.][]{DeMeo2014}. To explore possible solutions to this problem, we consider a more realistic disk -- one with a dead zone -- in following Section. 

\begin{figure}
  \centering
    \includegraphics[width=8.3cm]{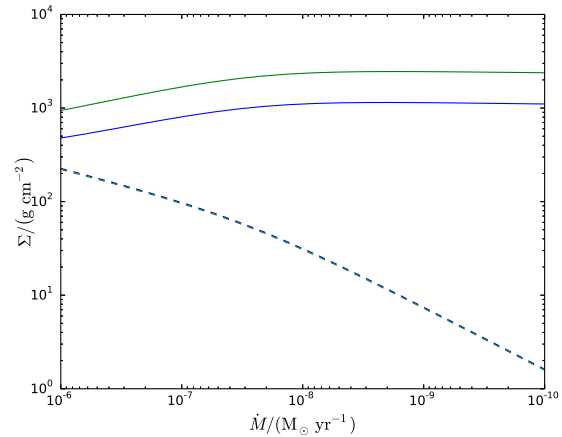}
     \caption{Surface density at the snow line radius, $R = R_{\rm snow}$, in the steady state disk for the nitrogen snow line radius (blue lines) and the ammonia snow line radius (green lines). The solid lines are disks with self-gravitating dead zone and dashed lines, which lie on top of each other, are fully MRI turbulent disks with $\alpha_{\rm m}=0.01$.}
     \label{sld}
\end{figure}

The dashed lines in Fig.~\ref{sld} show the surface density at the snow line radius as a function of the accretion rate. For low accretion rates, the surface density at the snow line radius becomes very small, $\Sigma \lesssim 10\,\rm g\, cm^{-2}$. Assuming a gas to solid ratio of 100 \citep[e.g.][]{bohlin1978,Hayashi1981}, the surface density of solid material is $\lesssim 0.1\,\rm g\, cm^{-2}$. In the fully turbulent disk model there is little solid material available for planetesimal formation in the outer parts of the disk.

\subsection{A Disk With A Dead Zone}

The MRI requires a sufficiently high level of ionization  in the disk for it to operate. The inner regions of a protoplanetary disk are hot enough to be thermally ionized where $T_{\rm c}\geq T_{\rm crit}$. The critical temperature, $T_{\rm crit}$, is thought to be around $800\,$K for activating the MRI \citep{Umebayashi1988}. However, farther from the host star, external sources such as cosmic rays and X-rays from the central star ionize only the surface layers up to a critical surface density, $\Sigma_{\rm crit}$. The value of the critical surface density is not well determined \citep{Martinetal2012a,Martinetal2012b}. For T Tauri stars, cosmic rays ionize a critical surface density of $\Sigma_{\rm crit} \approx\, 100\, \rm g\, cm^{2}$ \citep{Gammie1996}. However, this value could be much lower if we consider the chemical reaction of charged particles \citep{Sano2000}.  \citet{Bai2009} and \cite{Turner2008} found  that the MRI could operate in the surface layer of a disk ionized  by stellar X-rays with $\Sigma_{\rm crit} \approx 1 - 30\, \rm g\, cm^{-2}$.  Where the surface density is larger than this, $\Sigma \geq \Sigma_{\rm crit}$, a dead zone forms at the disk mid-plane where the MRI cannot operate \citep[e.g.][]{Gammie1996, Gammie1998}. 

In the dead zone layer, material builds up due to the lower viscosity and the small mass transport rate. The disk becomes self-gravitating when the \cite{Toomre1964} parameter, $Q=c_{\rm s}\,\Omega/\,\pi G \Sigma$ is less than the critical, $Q_{\rm crit}=2$. Then, a second type of turbulence, gravitational turbulence, is driven, with a viscosity given by
\begin{equation}
\nu_{\rm g} = \alpha_{\rm g} \frac{c_{\rm g}^2}{\Omega}
\end{equation}
and the alpha parameter is 
\begin{equation}
\alpha_{\rm g} = \alpha_{\rm m}\exp(-Q^4)
\end{equation}
\citep[e.g.][]{Zhuetal2010a}. The disk can reach a steady state with a self--gravitating dead zone \citep[e.g.][]{MartinandLubow2013prop,MartinandLubow2013dza,Rafikov2015}. \citet{MartinandLivio2012} found that the evolution of the snow line is significantly altered in a time--dependent  disk with a dead zone compared with a fully turbulent disk model. The build up of material can lead to the disk being gravo-magneto unstable \citep[see the disk structure sketches in ][]{MartinandLivio2013snow}. The temperature in the dead zone increases sufficiently for the MRI to be triggered there, leading to an accretion outburst. However, the outer parts of the disk that we are interested in here remain in a steady state. Thus, we consider only steady state self-gravitating disk solutions in this work.

We follow  \citet{MartinandLivio2012}  to obtain an analytic solution for the snow line radii assuming that both the ammonia and nitrogen snow lines are in the self--gravitating part of the dead zone. The disk has MRI active layers over a self-gravitating dead zone but we assume that $\Sigma \gg \Sigma_{\rm crit}\approx 0$. The self-gravitating disk has a surface density 
\begin{equation}
\Sigma=\frac{c_{\rm g}\Omega}{\pi G Q}.
\end{equation}
The accretion rate of the steady disk is 
\begin{equation}
\dot{M}=3\,\pi\, \nu_{\rm g}\,\Sigma=\left( \frac{3c_{\rm g}^3 \alpha_{\rm m}}{G} \right)\frac{\exp(-Q^4)}{Q}  .
\label{mdot-dz}
\end{equation}
The term in brackets is constant for a fixed snow line temperature. This expression depends sensitively on $Q$ but $Q$ is approximately constant in the range of accretion rates we use. We scale the variables to $T'_{\rm N_2, snow}= T_{\rm N_2, snow}/58 {\rm K}$, $T'_{\rm NH_3, snow}= T_{\rm NH_3,snow}/131 \, {\rm K}$, $\alpha'_{\rm m}=\alpha_{\rm m}/0.01$, $M'_{\star}=M_{\star}/\rm M_{\odot}$, $R'_\star=R_{\star}/3\,\rm R_{\odot}$ and $R'=R/\rm au$. Then, we can solve equation~(\ref{mdot-dz}) to calculate the scaled Toomre parameter
\begin{equation}
Q'=\frac{Q}{Q_{\rm crit}}=0.69\left[\frac{W(x)}{W(x_{0})} \right]^{\frac{1}{4}},
\end{equation}
where we define
\begin{equation}
x_{\rm N_{2}}=x_{\rm 0,N_{2}}\frac{\alpha_{\rm m}^{'4}T_{\rm N_{2},snow}^{'6} }{M_{\star}^{'4}}
\end{equation}
with $x_{\rm 0,N_{2}}\ =3.86\times10^{10}$  and
\begin{equation}
x_{\rm NH_{3}}=x_{0,\rm NH_{3}}\frac{\alpha_{\rm m}^{'4}T_{\rm NH_{3},snow}^{'6} }{M_{\star}^{'4}}
\end{equation}
with $x_{0,\rm NH_{3}}=5.13\times10^{12}$.
The equation 
\begin{equation}
x = W(x)\exp[W(x)]
\end{equation}
defines the Lambert function, $W$.

The steady state energy equation of a disk with a self--gravitating dead zone is  
\begin{equation}
\sigma T_{\rm e}^4 = \frac{9}{8}\nu_{\rm g}\Sigma\Omega^2+\sigma T_{\rm irr}^4.
\label{Tedz}
\end{equation}
The mid--plane temperature of the disk is related to the disk surface temperature  with equations~(\ref{Trr})--(\ref{opacity}). In Fig.~\ref{sla}, the two solid lines show the snow line radii that are the solutions to $T_{\rm c}=T_{\rm snow}$ with equation~(\ref{Tedz}). The snow line radius in the disk model with a dead zone is farther from the central star than in the fully turbulent disk model since the temperature of the disk is higher.  With a dead zone, for low accretion rates the snow line radius is insensitive to the accretion rate. The self--gravitating dead zone prevents the snow lines from moving too close to the central star in the late stages of disk evolution.

For lower accretion rates ($\dot{M}\leq 10^{-8}\, \rm M_{\odot}\,yr^{-1}$), irradiation dominates over the viscous heating term and we approximate equation~(\ref{Tedz}) with $T_{\rm e}=T_{\rm irr}$ and find the analytic snow line radius for nitrogen to be
\begin{equation}
R_{\rm N_{2}, snow} = 12.6\,{M}_\star^{'\frac{1}{9}}R_\star^{'\frac{2}{3}}T_\star^{'\frac{8}{9}} T_{\rm N_{2}, snow}^{'-0.31}\left[\frac{W(x)}{W(x_{0})} \right]^{-\frac{1}{18}}\, \rm au
\end{equation}
and for ammonia 
\begin{equation}
R_{\rm NH_{3}, snow} = 9.8\,M_\star^{'\frac{1}{9}}R_\star^{'\frac{2}{3}}T_\star^{'\frac{8}{9}} T_{\rm NH_{3}, snow}^{'-0.31}\left[\frac{W(x)}{W(x_{0})} \right]^{-\frac{1}{18}} \rm au.
\end{equation}
These values are given approximately by 
\begin{align}
R_{\rm N_{2}, snow} = & \,\,12.6\,\left(\frac{M_{_\star}}{\rm M_{\odot}}\right)^{\frac{1}{9}}\left(\frac{R_{_\star}}{3\ \rm R_{\odot}} \right)^{\frac{2}{3}}\left(\frac{T_{_\star}}{\rm 4000K} \right)^{\frac{8}{9}}\cr
& \times\left(\frac{T_{\rm N_{2}, snow}}{\rm 58K} \right)^{-0.31} \,\rm au
\label{dzsnowline1}
\end{align}
and
\begin{align}
R_{\rm NH_{3}, snow} = & \,\, 9.8\,\left(\frac{M_{_\star}}{\rm M_{\odot}}\right)^{\frac{1}{9}}\left(\frac{R_{_\star}}{3\ \rm R_{\odot}} \right)^{\frac{2}{3}}\left(\frac{T_{_\star}}{\rm 4000K} \right)^{\frac{8}{9}} \cr & \times \left(\frac{T_{\rm NH_{3}, snow}}{\rm 131K} \right)^{-0.31} \, \rm au.
\label{dzsnowline2}
\end{align}
The numerical values of equations~(\ref{dzsnowline1}) and~(\ref{dzsnowline2}) for our standard parameters are consistent with the solid lines in Fig.~\ref{sla} at low accretion rates.

The solid lines in Fig.~\ref{sld} show the surface density of the disk with a self--gravitating dead zone at the snow line radius, for nitrogen (blue line) and for ammonia (green line). The disk with the dead zone has a roughly constant surface density with accretion rate, particularly for low accretion rates. The surface density is much higher than even the maximum possible value for the critical surface density that may be MRI active,  $\Sigma_{\rm crit} \approx 200\,\rm g\, cm^{-2}$. Thus, our assumption of $\Sigma \gg \Sigma_{\rm crit}$ is justified. Furthermore, the surface density at the snow line radii is significantly higher in a disk with a dead zone thus allowing for more solid bodies to form.

For the low accretion rates expected at the end of the disk lifetime, the snow line radii for nitrogen and ammonia in the solar system are both farther from the Sun than the main asteroid belt. The ammonia snow line lies at about $9\,\rm au$ while the nitrogen snow line is at $12\,\rm au$. Thus, comets in the Kuiper belt are expected to form with significantly more nitrogen than asteroids in the main asteroid belt. In the next Section we consider observational constraints from the isotopic ratio of nitrogen, and estimate the fraction of the present Earth's atmosphere that could have been delivered by comets.

\section{The delivery of cometary nitrogen to Earth}
\label{nitrogen}
   
For the Earth's atmosphere, the present ratio of $^{15}$N/$^{14}$N is the combined result of the isotopic ratio of the primitive Earth's atmosphere, of asteroids and of comets. We follow the model of \citet{Hutsemekers2009} to find the percentage of nitrogen that may have been brought to Earth by comets. 

The isotopic ratio in the present atmosphere is measured to be ($^{15}$N/$^{14}$N$)_{\rm t}= 3.676\,\times\,10^{-3}$ \citep{JUNK1958}.  In contrast, the isotopic ratio of the primitive Earth, $(^{15}$N/$^{14}$N$)_{\rm p}$, is uncertain. Based on the value of the ratio in enstatite chondrite-type material and measurement of the Earth's interior, we shall therefore consider the isotopic ratio of the primitive Earth's atmosphere $(^{15}$N/$^{14}$N$)_{\rm p}$, to be $3.55\,\times\,10^{-3}$ which reflects the ingredients of the Earth in the protosolar nebula \citep{Cartigny1998,Cartigny2013,javoy1986,Jia2004,Marty1999,Murty1997}.

The nitrogen isotopic ratio of the primitive Earth is very similar to that observed in asteroids, thus we consider the nitrogen isotopic ratio of the present Earth as a mixture of that from comets and that from the primitive Earth's atmosphere. The abundances of each isotope are equal to the sum of that in comets and that in the primitive Earth so that 
\begin{align}
n_{\rm p}(^{14}{\rm N})+n_{\rm c}(^{14}{\rm N})=n_{\rm t}(^{14}\rm N) \label{NNpratio} 
\end{align}
and
\begin{align}
n_{\rm p}(^{15}{\rm N})+n_{\rm c}(^{15}{\rm N})=n_{\rm t}(^{15}\rm N),
\end{align}
where $n$ is the abundance and the subscripts p, c and t represent the primitive Earth, comets and present Earth, respectively. The terrestrial isotopic ratio weighted by the relative abundances of nitrogen from comets and from the primitive Earth is
\begin{equation}
(^{15}\rm{N}/^{14}\rm{N})_{\rm{t}}=\frac{n_{\rm c}(^{14}N)}{n_{\rm t}(^{14}N)}(^{15}\rm{N}/^{14}\rm{N})_{\rm c} + \frac{n_{\rm p}(^{14}N)}{n_{\rm t}(^{14}N)}(^{15}\rm{N}/^{14}\rm{N})_{\rm p}.
\end{equation}
Combining this with equation~(\ref{NNpratio}) gives the fraction of the abundance of $^{14}$N in comets to the total abundance
\begin{equation}
\frac{n_{\rm c}(^{14}\rm N)}{n_{\rm t}(^{14}\rm N)}=\frac{(^{15}\rm{N}/^{14}\rm{N})_{\rm p}-(^{15}\rm{N}/^{14}\rm{N})_{\rm t}}{(^{15}\rm{N}/^{14}\rm{N})_{\rm p}-(^{15}\rm{N}/^{14}\rm{N})_{\rm c}}.
\label{ratio}
\end{equation}
This is the fraction of the nitrogen that was delivered to the Earth by comets compared to the total amount of nitrogen in the atmosphere
\citep{Hutsemekers2009}. 

The isotopic ratio of nitrogen in comets is uncertain. \citet{McKeegan2006} analyzed COMET 81P/Wild2 grains returned by Stardust and found a bulk isotopic ratio that is close to the terrestrial and chondritic value $(^{15}$N/$^{14}$N$)_{\rm c}=\,3.55\,\times\,10^{-3}$ but most comets have a higher value. For comets measured remotely using optical-UV spectroscopy of CN and radio observations of HCN, the isotopic ratio of nitrogen was found to be in the range of $(^{15}$N/$^{14}$N$)_{\rm c} = 4.87 - 7.19 \,\times\,10^{-3}$ \citep{Arpigny2003,Bockelee2008a,Bockelee2008b,Hutsemekers2005,Jehin2004,Jehin2006,Jehin2008,manfroid2009}. In our models, we consider $(^{15}$N/$^{14}$N$)_{\rm c}$ values in the range of  $5.00 - 7.00\,\times\,10^{-3}$.  

\begin{figure}
  \centering
    \includegraphics[width=8.3cm]{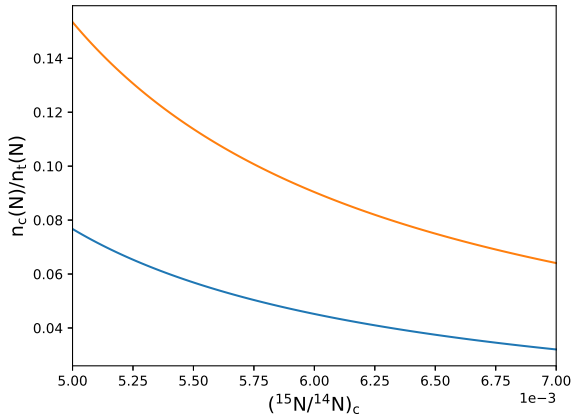}
     \caption{The fraction of the present Earth's nitrogen that was delivered by comets, $n_{\rm c}({\rm N})/n_{\rm t}({\rm N})$,  as a function of the cometary isotopic nitrogen ratio, $(^{15}$N/$^{14}$N$)_{\rm c}$. 
The blue line is the model with the primitive Earth ratio of $(^{15}$N/$^{14}$N$)_{\rm p}=\,3.676\,\times\,10^{-3}$ and the yellow line is the model with $(^{15}$N/$^{14}$N$)_{\rm p}=\,3.786\,\times\,10^{-3}$.}
     \label{NNratio}
\end{figure}

The current isotopic ratio of nitrogen in the Earth's atmosphere is $(^{15}$N/$^{14}$N$)_{\rm t}=\,3.676\,\times\,10^{-3}$. The blue line in Fig.~\ref{NNratio} shows that about 3 - 8 \% of the Earth's atmospheric nitrogen was delivered by comets if no other processes have changed the isotopic ratio since the comets were delivered. However, this current day isotopic ratio may be lower that of the Archaean atmosphere which reflects the effect of the late heavy bombardment (LHB). Nitrogen content of the Archaean atmosphere was brought to subduction zones \citep{mallik2018} and it results in the current isotopic ratio of nitrogen in the Earth's atmosphere.

If we consider the higher isotopic ratio of nitrogen in Archean Earth's atmosphere of  $(^{15}$N/$^{14}$N$)_{\rm t}=\,3.786\,\times\,10^{-3}$ \citep{Jia2004}, the yellow line in Fig.~\ref{NNratio} shows that about 6 - 15\% of the Earth's atmospheric nitrogen was delivered by comets. 

In summary, if the isotopic ratio of nitrogen has not changed since the Archean era, the cometary contribution of nitrogen is only about 5\%. On the other hand, if in the Archean era, the Earth has an enriched $^{15}$N atmosphere, the cometary contribution of nitrogen may be about 10 \%.

\section{Debris disk evolution}
\label{asteroid}

\begin{figure*}\centering
    \includegraphics[width=17.0cm]{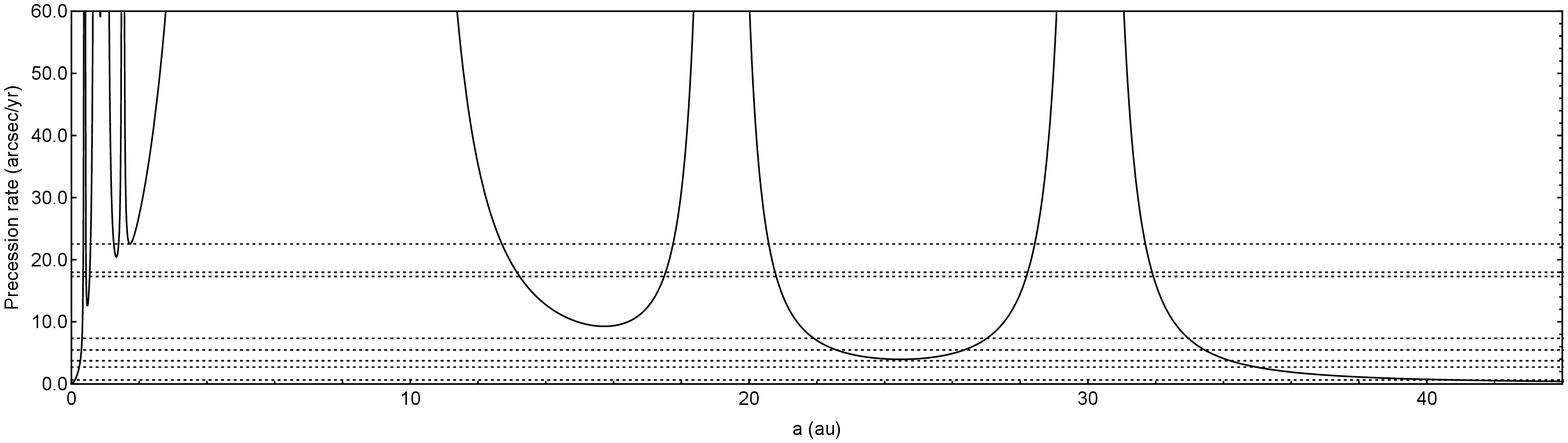}
    \includegraphics[width=17.0cm]{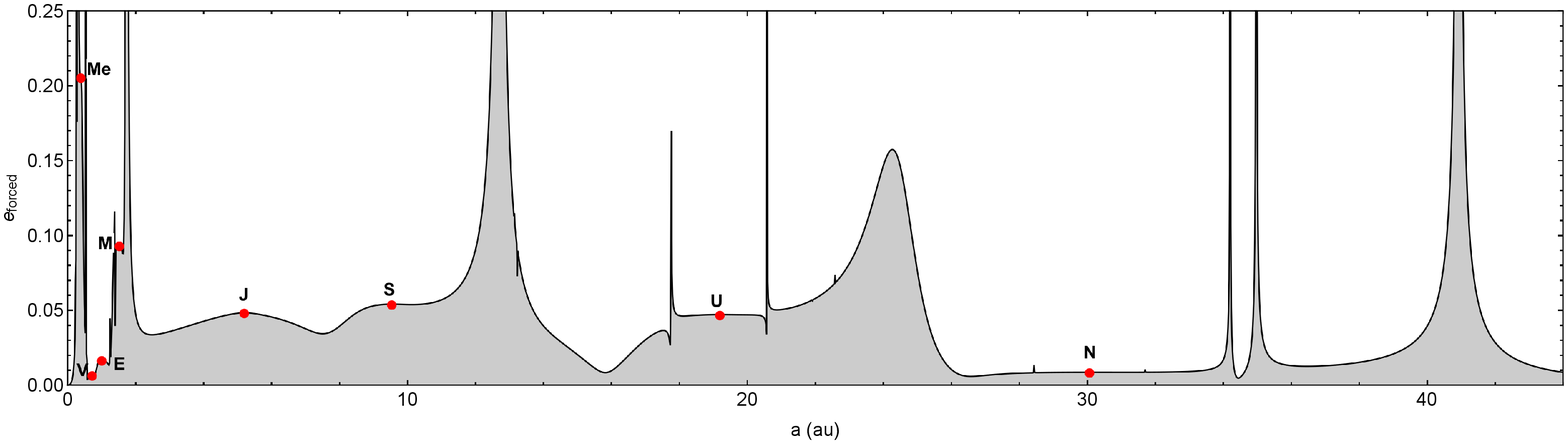}
     \caption{Top panel: The apsidal precession rate of a test particle as a function of semi--major axis (solid lines) and the eigenfrequencies for the planets (horizontal-dotted lines) (see Table~\ref{eigen} for specific values). The intersection of the test particle's precession rate with a planetary eigenfrequency represents the location of a apsidal secular resonance. Bottom panel: The maximum forced eccentricity of a test particle as a function of semi--major axis. The wider the gray region, the more asteroids/comets will undergo resonant perturbations.}
     \label{sec}
\end{figure*}

We now consider possible delivery mechanisms for nitrogen containing comets in the Kuiper belt to collide with Earth.  We first consider an analytic secular resonance model and then we use N--body simulations to model the delivery of comets to the Earth.

\subsection{Resonances in the Kuiper belt  }

In the context of the Kuiper belt, interior mean-motion resonances with Neptune generally produce scattering events from eccentricity excitations compared with exterior mean-motion resonances that provide resonant capture of debris \citep{Beauge1994}. Thus, we narrow our investigation of a nitrogen transport mechanism to secular resonances rather than including mean-motion resonances, since there are no interior mean-motion resonances within the Kuiper belt.
There are two types of secular resonances, apsidal and nodal. An apsidal resonance excites eccentricity, while a nodal resonance excites inclination \citep{Froeschle1991}. Apsidal resonances are the most important secular resonances for transporting material to the Earth, since the comets remain in the orbital plane of the  system's solid large bodies with an increased eccentricity and are more likely to collide with the Earth.  Here, we apply a first-order secular resonance model to the solar system to find all of the the apsidal resonance locations. We are particularly interested in secular resonances that are outside of the ammonia snow line (at around $9\,\rm au$ at the end of the gas disk lifetime, see Section~\ref{disk}) since objects  outside of this location may contain large amounts of nitrogen.

%\subsection{Analytical Secular Resonance Model}

A secular resonance arises when the free precession rate of a test particle is equal to the eigenfrequency of one of the planets \citep[e.g.][]{Froeschle1986,Yoshikawa1987,Minton2011,Smallwood2018a,Smallwood2018b}.  The planetary eigenfrequencies are obtained by calculating the eigenvalues from the Laplace-Lagrange equations in the context of the generalized secular perturbation theory \citep[see for example][]{murray1999book}. The eigenfrequencies for each of the solar system planets are shown in Table~\ref{eigen}.  The free precession rate of a test particle with orbital frequency, $n$, and semi--major axis, $a$, in the potential of the solar system  is
\begin{equation}
\dot{\omega} = \frac{n}{4}\sum_{j=1}^N\frac{m_j}{m_*}\alpha_j \bar{\alpha}_j b_{3/2}^{(1)}(\alpha_j),
\end{equation}
\citep{murray1999book} where $m_j$ are the masses of the solar system planets, $m_*$ is the mass of the Sun, $b^{(j)}_{s}$ is the Laplace coefficient and the coefficients $\alpha_j$ and $\bar{\alpha}_j$ are defined as
\begin{equation}
  \alpha_{j}=\begin{cases}
    a_j/a, & \text{if $a_j < a$},\\
    a/a_j, & \text{if $a_j > a$},
  \end{cases}
\end{equation}
and
\begin{equation}
  \bar{\alpha}_{j}=\begin{cases}
    1, & \text{if $a_j < a$},\\
    a/a_j, & \text{if $a_j > a$}.
  \end{cases}
\end{equation}
The top panel of Fig.~\ref{sec} shows the eigenfrequencies of the planets as horizontal dashed lines. The test particle free precession rate is represented by the solid lines. Where the solid lines cross a horizontal line there is an apsidal secular resonance.

\begin{table}
	\centering
	\caption{The apsidal eigenfrequencies for each of the solar system planets.}
	\label{tab:example_table}
	\begin{tabular}{lc} % four columns, alignment for each
		\hline
		Planet & Apsidal Eigenfrequency ($g_i$)\\
        & (arcsec/yr) \\
		\hline
		Mercury (Me)  & $5.46347$ \\
        Venus (V)  & $7.35054$  \\
        Earth (E)  & $17.2968$  \\
        Mars (M)  & $17.9887$  \\
        Jupiter (J)  & $3.72802$ \\
        Saturn (S)  & $22.5375$  \\
	    Uranus (U)  & $2.70839$   \\
        Neptune (N)  & $0.633873$  \\
		\hline
	\end{tabular}
    \label{eigen}
\end{table} 

We now consider the strength of the apsidal resonances. In the bottom panel of Fig.~\ref{sec} we show the forced eccentricity of a test particle in the solar system  (for details on these calculations see \cite{murray1999book} and \cite{Minton2011}). Planetary debris that falls within the high eccentricity parts of the gray-shaded regions undergoes eccentricity growth that leads to ejection or collision with a larger object. In the context of nitrogen transport to Earth, the source of nitrogen may originate from the Kuiper belt which extends roughly from $30\, \rm au$ to $50\, \rm au$ \citep{Duncan1995}. The $\nu_8$ secular resonance is the most prominent secular resonance within the Kuiper belt \citep{Froeschle1986}. It is produced when the free precession rate of the comets is close to the eigenfrequency of Neptune. According to our linear secular resonance model, the $\nu_8$ resonance is located at $40.9\,\rm au$ (the resonance location is based on first-order approximations) which is consistent with the actual location of the resonance \citep{Nagasawa2000}. 

Populations of Kuiper belt objects that are in resonance suggest that Uranus and Neptune underwent outward migration \citep{Fernandez1984,Malhotra1993,Malhotra1995,Hahn1999,Tsiganis2005,Hahn2005,Chiang2002,Chiang2003,MurrayClay2005,Morbidelli2007,Nesvorny2015,Nesvorny2016}. During this dynamical phase, resonant bodies in the Kuiper belt were destabilized and there was a sudden massive delivery of planetesimals to the inner solar system \citep[e.g.][]{Levison2003,Gomes2005,Nesvorny2017}.  In particular, the $\nu_8$ secular resonance swept through the inner parts of the Kuiper belt \citep{Nagasawa2000} and caused a large fraction of the comets to acquire free precession frequencies similar to that of Neptune. Comets that originated or got "bumped" into this sweeping resonance would have undergone increased eccentricity excitations leading to collisions with Earth, allowing for the transport of nitrogen.

\subsection{N--body simulations of Earth impacts} 
To study the delivery of comets to the Earth, we use the $N$-body simulation package, REBOUND, to model the Kuiper belt including the Earth, the four giant planets and Sun in the solar system \citep{Rein2012b}. We use a hybrid integrator called $mercurius$. This integrator uses WHFast which is an implementation of a Wisdom-Holman symplectic integrator to simulation the long-term orbit integrations of the planetary system. For close encounters of test particles and planets, it switches smoothly to IAS15 which is a 15th-order integrator to simulate gravitational dynamics \citep{Rein2015a, 2015MNRAS.452..376R}. Thus, we can calculate the evolution of the Kuiper belt for a duration of one hundred million years (Myr) quickly and precisely. In our simulation, we take a snapshot every 0.1 Myr.

Comets interact with the Earth, the gas giant planets and the Sun due to gravitational forces while there is no interaction between  comets. The interaction time scale of some of the largest asteroids are of the order of magnitude of the age of the Solar System \citep{Dohnanyi1969}.

The obit of each comet is defined initially by six orbital elements, $a,i,e,\omega,\Omega,\nu$. In our model, we assume the semi--major axis, $a$, is distributed uniformly in the range $a_{\rm min}$ = 38 to $a_{\rm max}$ = 45 au including the region of the $\nu_8$ secular resonance (see Fig.~\ref{sec}). The uniform distribution is given by 
\begin{equation}
a_{\rm comet} = (a_{\rm max}-a_{\rm min})\,\times\,\chi_{r}+a_{\rm min}
\end{equation}
\citep{Lecar1997,Smallwood2018a} and $\chi_{\rm r}$ is a randomly generated number between 0 and 1.
The inclination, $i$, is randomly allocated in the range from 0$^\circ$ to 10$^\circ$ and the eccentricity, $e$, is randomly allocated in the range from 0 to 0.1. The argument of periapsis, $\omega$, the longitude of the ascending node, $\Omega$ and the true anomaly, $\nu$ are all uniformly distributed in the range from 0$^\circ$ to 360$^\circ$.

The present-day orbital elements for each of the planets are set as the initial parameters for the planets because the Solar system is stable over long time-scales \citep{Duncan1998,Ito2006}. REBOUND queries those data automatically from NASA HORIZONS database. A test particle that has a semi--major axis beyond $200\,\rm au$ is considered to be ejected. 

In our simulation we inflate the size of the Earth to its Hill radius. The Hill radius is given by 
\begin{equation}
R_{\rm H} = a_{\rm p} \left(  \frac{M_{\rm planet}}{3 \rm M_{\odot}}  \right)^{1/3},
\end{equation}
where $M_{\rm planet}$ is the mass of the planet and $a_{\rm p}$ is the semi--major axis of the planet \citep{Delsanti2006}. The Hill radius for the Earth is $2 \times 10^6\,$km. In principle, three Hill radii is the range of the planet's gravitational force and comets cannot remain on stable orbits within this region \citep{Gladman1993,Chatterjee2008,Morrison2015}. The increased size of the Earth  artificially increases the rate of collisions with the Earth allowing us to simulate fewer particles than would have been present in the early Kuiper belt. 

\begin{figure}
  \centering
    \includegraphics[width=8.3cm]{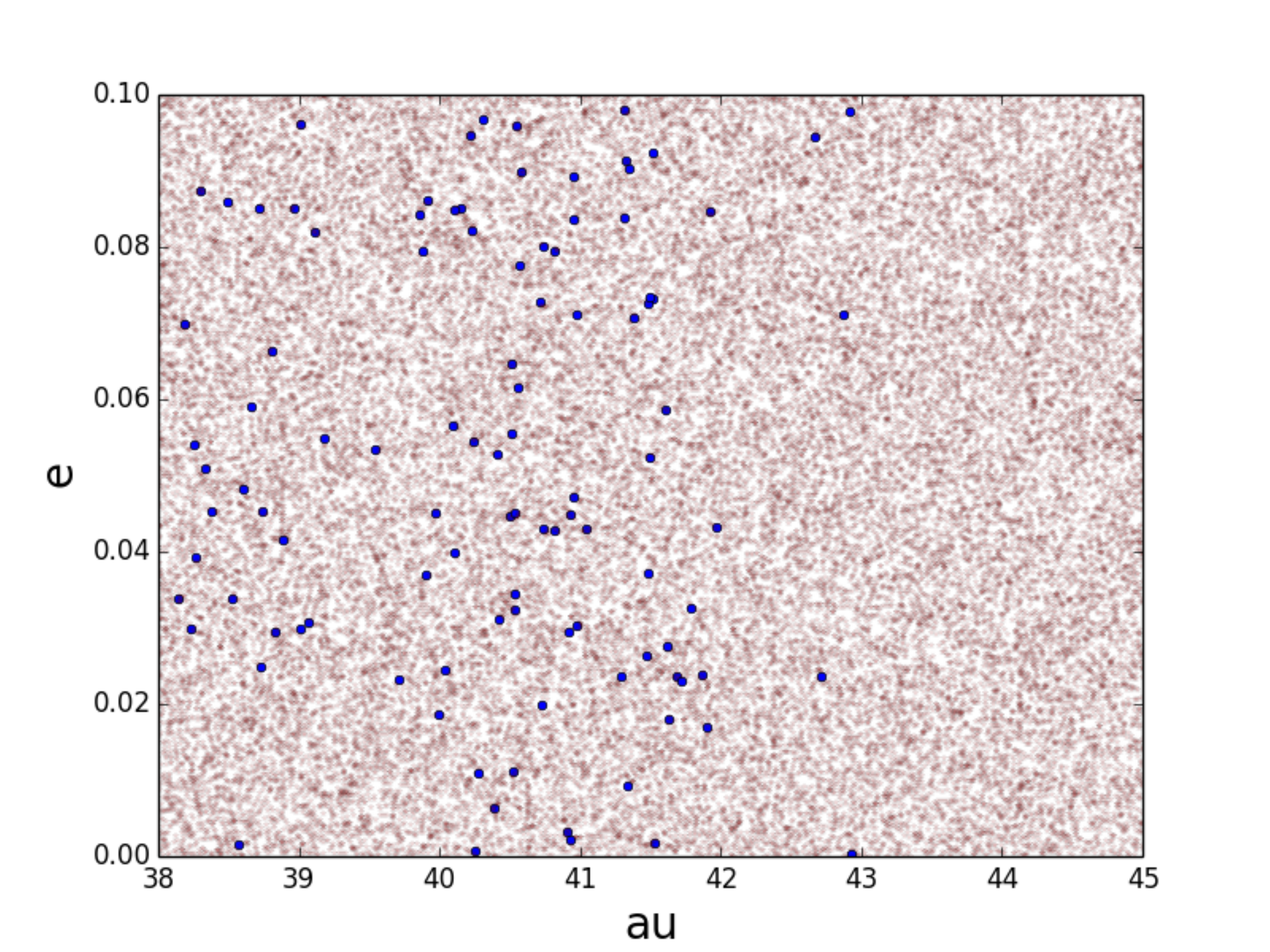}
     \caption{Initial distribution of semi--major axis and eccentricity of 50,000 test particles. Blue dots are particles that collide with the Earth.}
     \label{cearth}
\end{figure}

Fig.~\ref{cearth} shows the initial eccentricity and semi--major axis distribution of 50,000 test particles.  There are a total of 104 test particles that collide with the Earth within 100 Myr and the initial conditions for these are shown by the blue dots. 70 of these test particles originate from the region of the $\nu_{8}$ resonance around $40-42\,\rm au$. The 3:2 and 5:3 mean motion resonances of Neptune that are at 39.5 and $42\,$au also contribute a few of the collisions with the Earth. 

\begin{figure}
  \centering
    \includegraphics[width=8.3cm]{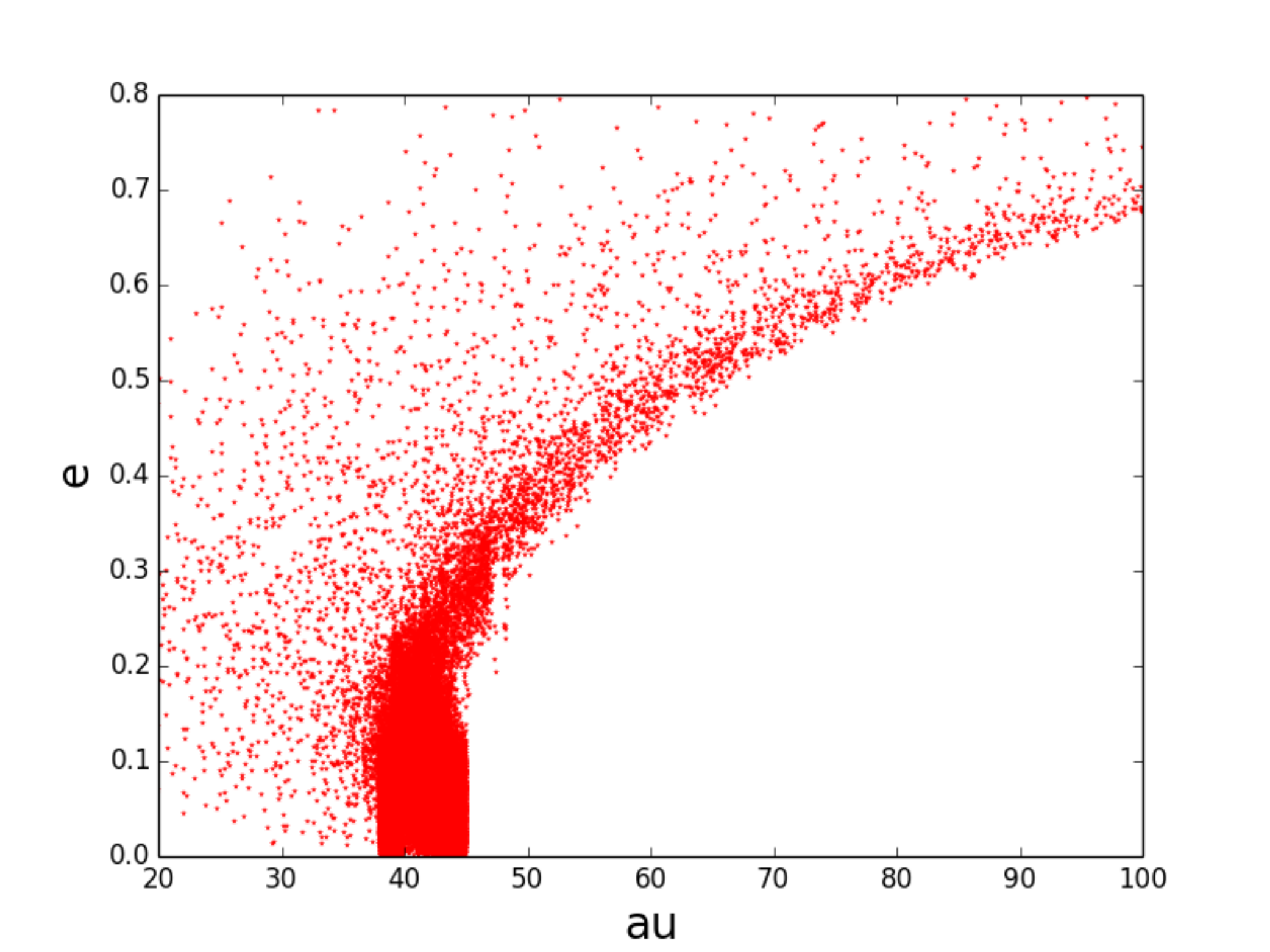}
     \caption{Final distribution of semi--major axis and eccentricity of the test particles that remain in the simulation.}
     \label{fearth}
\end{figure}

On the other hand, a test particle that has semi--major axis $a > 43\,$au is quite stable and the region is regarded as the classic Kuiper belt objects and the final distribution of eccentricity and semi--major axis of the comets in fig.~\ref{fearth} is roughly consistent with that observed scattered Kuiper belt objects. However, there are still many particles in the inner region where $a < 40\,$au. Current observations show that this region is quite clear  \citep[see fig. 7 in][]{jewitt2009}. Since we do not consider the outward migration of Uranus and Neptune in our simulation, we do not reproduce this clearing.  The region of the $\nu_8$ resonance is not completely cleared out of comets yet. About 50\% of the test particles that were initially distributed in the region of the $\nu_8$ resonance are still in this region. %Hence, we expect the collisions to continue for 100 Myr.

%However, we do not consider the outward migration of Uranus and Neptune, the region where $a > 40\,$au is still stable in our simulation which is somewhat different  to current observations \citep[see fig. 7 in][]{jewitt2009}.  

%{\bf it's not clear to me what is different and what is the same, please try to explain. Do you mean the region 40--43 is different?}

%To study the orbital evolution of the particles that impact the Earth, Fig. \ref{trace} shows 104 semi--major axes of test particles with time. 

%\begin{figure}
%  \centering
%    \includegraphics[width=8.3cm]{au.eps}
%     \caption{Semi--major axes--time graph and eccentricities--time graph of 104 impact particles.}
%     \label{trace}
%\end{figure}

\begin{figure}
  \centering
    \includegraphics[width=8.3cm]{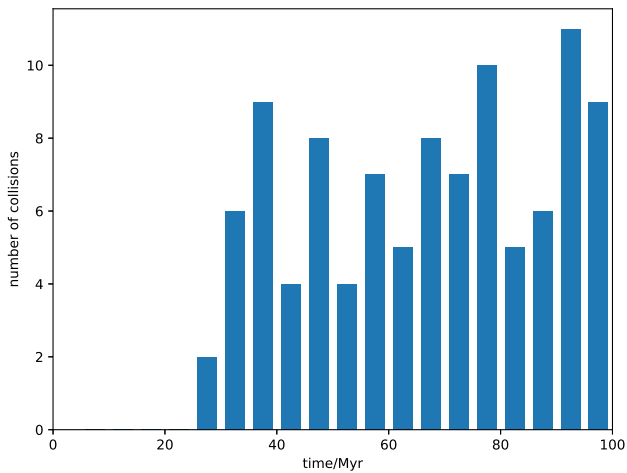}
     \caption{Histogram of the number of collisions of comets with the Earth as a funciton of time. There are 104 impact particles within 100 Myr.}
     \label{hitrace}
\end{figure}

Resonances in the Kuiper belt excite the eccentricities of the test particles and this increases the chance of a close encounter with a planet. These close encounters result in outward or inward migration of test particles. Once a test particle comes in to the inner solar system, the Earth has an opportunity to capture it. Fig.~\ref{hitrace} shows that particles begin colliding with the Earth after about 25 Myr and there are more than 4 impact particles every 5 Myr years after 30 Myr. In the next section we use the results of our N--body simulation to estimate the amount of nitrogen that was delivered to the Earth from the Kuiper belt. 

\subsection{Nitrogen delivery to the Earth}

We now consider a simple calculation to scale our simulation result to calculate how much mass may have been delivered to the Earth by comets. Most of the impacts come from the $\nu_8$ resonance region and we focus on these.
The fraction of comets that begin in the region we are interested in 40-42 au, assuming a uniform distribution of the number of test particle in semi--major axis. However, we inflated the size of the Earth to increase the possibility of a collision. \citet{Smallwood2018a} calculated that the number of collisions is reduced by a factor of 800/5 if we shrink the size of the Earth from its Hill's radius to the radius of the Earth. Thus, the collision fraction should be $3\times10^{-5}$ collisions per particle per $100\,$Myr.  

%Because we focus on the $\nu_{8}$ region (40 - 42 au), assuming a uniform distribution of mass in semi--major axis, the initial mass there is a fraction of $2/7$ of the initial mass of the Kuiper belt, about $1.7\times10^{28}$g. 
%The initial mass of the Kuiper belt was at least 10 times the mass of Earth, $\rm M_{\oplus}=6\times10^{27}$g \citep{Delsanti2006}. The amount of mass that originated in the $\nu_8$ resonance and collided with the Earth is approximately $ 5\times10^{23}$g in $100\,$Myr. Because at the end of our simulation there are still $50\%$ of the original particles in the region of the $\nu_8$ resonance, we multiply the number of collisions by a factor of 2 to find the mass to be $\approx 1 \times 10^{24}\,$g. For comparison, the total mass that was delivered to the Earth during LHB is estimated to be around $2\times 10^{23}-2\times 10^{24}$g by modeling the thermal effects of the LHB \citep{abramov2013}. The Moon only got about $9\times 10^{21}$g during the LHB and thus we can ignore the contribution of collisions to the Moon in our simulation \citep{Gomes2005}.

The initial mass of the Kuiper belt was at least 10 times the mass of Earth, $\rm M_{\oplus}=6\times10^{27}$g \citep{Delsanti2006}. The amount of mass that originated in the $\nu_8$ resonance and collided with the Earth is approximately $ 5\times10^{23}$g in $100\,$Myr. Because at the end of our simulation there are still $50\%$ of the original particles in the region of the $\nu_8$ resonance, we multiply the number of collisions by a factor of 2 to find the mass to be $\approx 1 \times 10^{24}\,$g. For comparison, the total mass that was delivered to the Earth during LHB is estimated to be around $2\times 10^{23}-2\times 10^{24}$g by modeling the thermal effects of the LHB \citep{abramov2013}. However, according to the constraint of  atmospheric $^{36}$Ar, \citet{Marty2016} estimated the cometary contribution to be about $2.0 \times  10^{22}\,$g.
This value is 2 orders of the magnitude lower than our estimate. However, we have not taken into account the fact that a comet starts to sublime volatile when it enters the inner solar system. For example, the dynamical lifetime of comet Halley is about 3000 perihelion passages or $0.2\,$Myr \citep{olsson1988}. It has lost about  0.4\% of the total mass of the nucleus over the last 2,200 yr \citep{Schmude2010}. From observations, we have already found some small comets that have radius $< 0.4\,$km that have lost of most their mass \citep[see table 9.4 in][]{fernandez2006}. If we consider a comet that has a nucleus radius of $1\,$km, the bulk density is $0.5\,\rm g\, cm^{-3}$ and the perihelion distance is $1 \,\rm au$, the dynamical lifetime is less than 1000 orbits  \citep[see fig. 9.2 in][]{fernandez2006}. Due to the mass loss of comets, comets only possess 1 \% to 10 \% of their original mass when they collide with the Earth. Thus, the mass delivered by comets in our model is in the approximate range of $10^{22}$ to $10^{23}\,$g.

To calculate how much nitrogen could have been delivered in these comets, we use data of 16 comets from \citet{le2015,Paganini2012} to estimate the mean values of highest relative abundances of NH$_3$ and HCN to be 0.7\% and 0.28\% with respect to water. Besides, in a recent study, ROSINA measured the N$_2$/CO ratio by mass to be 0.570 \% on the comet 67P/Churyumov-Gerasimenko and CO in this comet is about 10\% with respect to water \citep{le2015}. Thus, we use 0.057 \% for the relative abundance of N$_2$ with respect to water. H$_2$O takes up 50\% of the composition of the comet by mass and thus the total mass of nitrogen from comets is estimated to be in the range of 3.9 $\times 10^{19}$ to 3.9 $\times 10^{20}$g.

The total mass of the Earth's atmosphere is $5.136 \times 10^{21}$g \citep{weast1989}. Nitrogen makes up 78 \% of the composition of the current Earth's atmosphere so that the mass of nitrogen in the atmosphere is $3.8 \times 10^{21}\, $g \citep{verniani1966}. If 10\% of nitrogen was delivered by comets (see Section ~\ref{nitrogen}) the mass of cometary nitrogen is $4 \times 10^{20}$g and our simulation is consistent with this value. 

Note also that here we only consider the contribution from the $\nu_{8}$ region. There would have been additional collisions as a result of mean-motion resonances of Neptune (see fig.~\ref{cearth}) and the migration of Uranus and Neptune which we did not include in our simulation.  In the Archaean era, the partial pressure of nitrogen was at least 1.4 $\sim$ 1.6 times higher than today's atmosphere. \citet{mallik2018} estimated this value by subtracting the net degassing flux of N from arcs and back-arc basins \citep{Hilton2002}, mid-ocean ridges and intraplate-settings \citep{SANO2001} from the influx of N at subduction zones today. Due to N recycling efficiency, some nitrogen in the Earth's atmosphere was brought to the subduction zones and the deep mantle \citep{mallik2018}.  However, there are some contradictory results that the partial pressure of the nitrogen in Archaean atmosphere is similar or less than the present atmosphere's based on different analyses of Archean samples \citep{AVICE2018,som2016}. Consequently, it leaves some uncertainties and our simulations cannot solve this problem.

\subsection{Water delivery to the Earth}
Because comets also contain a large amount of water, they deliver both nitrogen and water and the D/H ratio in the Earth may also be changed. Since the D/H ratios of comets are much higher than terrestrial water and enstatite chondrites do not contain large amount of water, it could be used to constrain the contribution of water from comet to the Earth and it should be less than about $10\%$ \citep{bockelee2004,Dauphas2000}. Since water takes up $50\%$ of a comet's composition, we estimate the amount of water that delivered from comets and this value is $5 \times 10^{21}\sim \ 5 \times 10^{22}$g. The total mass of the water on the Earth is $1.35 \times 10^{24}$g. Therefore, the water contribution from comets is just about $0.4\% \sim 4\%$. This value is more in agreement with \citet{Marty2016} that states the cometary contribution of water is $\leq$ 1 \%,whatever the reservoir considered.

%More nitrogen would be delivered until the $\nu_{8}$ resonance cleared most of small objects around 41 au. Because comets may lose a lot of volatile when they come into the inner solar system, it is hard to estimate the final amount of nitrogen inside comets when they fall onto the Earth. 

%Besides, NH$_3$ takes up 0.13 \% from Spectra of the Halley, Hartley-Good, Thiele and Borrelly comets \citep{wyckoff1991}.
\section{Conclusions} \label{con}
Although it is not the main source of nitrogen in the Earth's atmosphere, a significant fraction of the present Earth's nitrogen may have come from comets (about 10\%).  The present isotopic ratio of $^{15}$N$/^{14}$N in the Earth's atmosphere is a combination of the primitive Earth value and the value contained in asteroids/comets that hit the Earth. With steady state protoplanetary disk models that include a self--gravitating dead zone, we find that the ammonia and nitrogen snow line radii were around $9\,\rm au$ and $12\,\rm au$, respectively,  at the end of the disk's lifetime. Thus, comets in the Kuiper belt probably contain a large amount of nitrogen. Besides, the Kuiper belt provides an extremely cold environment where CO is frozen out and large enhancements of the isotopic ratio $^{15}$N$/^{14}$N occur there. Since the Earth's atmosphere might be enriched in $^{15}$N in the Archean era, the nitrogen in the Kuiper belt is a possible source.  

Apsidal secular resonances played an essential role in bringing comets to the Earth. The most prominent secular resonance in the Kuiper belt, the $\nu_8$ resonance, might have excited the eccentricity of the comets. In the early stages of the solar system evolution, Uranus and Neptune both underwent outward migration during which the $\nu_8$ resonance swept through comet-like objects from the inner Kuiper belt to its present location at around $41\,\rm au$. With numerical N--body simulations we have shown that comets around this region are scattered and may collide with the Earth when they enter the inner solar system. The period of the LHB corresponds with the early Archean era. During this time, the resonance most probably delivered comets and thus nitrogen to the Earth. With the supplement of the late delivery of nitrogen to the Earth, the isotopic ratio $^{15}$N$/^{14}$N of the Earth's atmosphere increased slightly. Our research quantifies the amount of nitrogen that is delivered by comets.

Exoplanetary systems that contain both a warm and a cold debris belt (like the solar system) may be common \citep[e.g][]{Moro-mart2010,Morales2011}. The outer debris disk is a potential source of nitrogen to terrestrial planets in the inner regions. For example, the planetary and debris disk system HR 8799 closely resembles that of the outer Solar system by having two massive debris disks that bracket four giant planets, with the outermost disk consisting of cold dust (T $\sim$ 45 K) \citep{Marois2008}. Investigating the delivery of nitrogen in such systems can contribute to the understanding of the atmospheric properties of exoplanets and planetary habitability.

\section*{Acknowledgments} 
All the simulations were run at the UNLV National Supercomputing Institute on the Cherry Creek cluster. CC acknowledges support from a UNLV graduate assistantship. JLS acknowledges support from a graduate fellowship from the Nevada Space Grant Consortium (NVSGC). RGM acknowledges support from NASA grant NNX17AB96G. Simulations in this paper made use of the REBOUND code which can be downloaded freely at http://github.com/hannorein/rebound.
\bibliography{cc.bib}

\end{document}